\def\etal{et~al.\ }
\def\Lya{{\rm Ly}\kern 0.1em$\alpha$}
\def\FeII{{\rm Fe}\kern 0.1em{\sc ii}}
\def\MgI{{\rm Mg}\kern 0.1em{\sc i}}
\def\MgII{{\rm Mg}\kern 0.1em{\sc ii}}
\def\MgIIwaves{{\rm Mg}\kern 0.1em{\sc ii}~$\lambda\lambda 2976, 2803$}
\def\CIV{{\rm C}\kern 0.1em{\sc iv}}
\def\SiIV{{\rm Si}\kern 0.1em{\sc iv}}
\def\NV{{\rm N}\kern 0.1em{\sc v}}
\def\OVI{{\rm O}\kern 0.1em{\sc vi}}
\def\CIVwaves{C\kern 0.1em{\sc iv}~$\lambda\lambda 1548, 1550$}
\def\HI{{\rm H}\kern 0.1em{\sc i}}
\def\HII{{\rm H}\kern 0.1em{\sc ii}}
\def\kms{\hbox{km~s$^{-1}$}}
\def\cm2{\hbox{cm$^{-2}$}}
\documentstyle[11pt,aaspp4,flushrt]{article}

\tighten
\begin{document}


\slugcomment{\it Accepted by the Astrophysical Journal}
\lefthead{Charlton \& Churchill}
\righthead{{\MgII} Absorbing Galaxies}


\title{The Kinematic Composition of {\MgII} Absorbers\altaffilmark{1}}

\thispagestyle{empty}

\author{\sc Jane~C.~Charlton\altaffilmark{2} 
        and 
        Christopher~W.~Churchill}
\affil{Astronomy and Astrophysics Department \\
       Pennsylvania State University,
       University Park, PA 16802 \\
       email: {\it charlton, cwc@astro.psu.edu}}

\begin{center}
Accepted for publication: {\it Astrophysical Journal}
\end{center}

\altaffiltext{1}{Based in part on observations obtained at the
W.~M. Keck Observatory, which is jointly operated by the University of
California and the California Institute of Technology.}
\altaffiltext{2}{Center for Gravitational Physics and Geometry,
                 Pennsylvania State University}


\begin{abstract}
The study of galaxy evolution using quasar absorption lines requires
an understanding of what components of galaxies and their surroundings
are contributing to the absorption in various transitions.
This paper considers the kinematic composition of the class of $0.4 <
z < 1.0$ {\MgII} absorbers, particularly addressing the question of what
fraction of this absorption is produced in halos and what fraction
arises from galaxy disks.
We design models with various fractional contributions from radial
infall of halo material and from a rotating thick disk component.  
We generate synthetic spectra from lines of sight through model
galaxies and compare the resulting ensembles of {\MgII} profiles with
the $0.4 \leq z \leq 1.0$ sample observed with HIRES/Keck.
We apply a battery of statistical tests and find that pure disk and
pure halo models can be ruled out, but that various models with
rotating disk and infall/halo contributions can produce an ensemble
that is nearly consistent with the data.
A discrepancy in all models that we considered requires the existence
of a kinematic component intermediate between halo and thick disk.
The variety of {\MgII} profiles can be explained by the gas in disks and
halos of galaxies not very much different than galaxies in the local
Universe.

In any one case there is considerable ambiguity in diagnosing the
kinematic composition of an absorber from the low ionization high
resolution spectra alone.  
Future data will allow galaxy morphologies, impact parameters, and
orientations, {\FeII}/{\MgII} of clouds, and the distribution of high
ionization gas to be incorporated into the kinematic analysis.
Combining all these data will permit a more accurate diagnosis of the
physical  conditions along the line of sight through the absorbing
galaxy.

\end{abstract}


\section{Introduction}
\label{sec:intro}

\pagestyle{myheadings}
\markboth{\sc Charlton \& Churchill \hfill Kinematic Models~~}
         {\sc Charlton \& Churchill \hfill Kinematic Models~~}

Quasar absorption lines (QALs) provide a diagnostic of the gaseous
conditions in and around galaxies as they form and evolve.
Most {\MgII} absorbers at $z < 1$ are produced within $40 h^{-1}$~kpc of
known $>0.1 L^*$ galaxies (\cite{ste95}).  
High resolution spectroscopy (HIRES/Keck) shows that they have
multiple absorbing components (\cite{cvc98}, hereafter CVC98).
These components should provide information about how the {\MgII} gas is
distributed spatially and kinematically.
More generally, multiple component structure is apparent in various
ionization stages of the different chemical elements probed by QALs
over the history of the Universe.  
Before QALs can be fruitfully applied to a {\it detailed} study of
conditions in galaxies we must resolve many of the remaining
ambiguities inherent in their interpretation.

Based upon low resolution spectra, the {\MgII}
absorbers with $W({\MgII}) > 0.3$~{\AA} have generally been interpreted
as material infalling into the halos of the normal $>0.1 L^*$ galaxies
(\cite{ste95}; \cite{mo96}).  
However, most $L^*$ galaxies are disk galaxies, and the disks
themselves contain the {\HI} necessary for ionization conditions
that allow {\MgII} to survive.   
Disks in the local Universe clearly extend well beyond the optical
radii (\cite {irw95}), but it is hard to establish directly the {\HI}
distribution at $N({\HI}) < 10^{19}${\cm2}. 
In a few cases, sensitive 21~cm measurements provide maps of galaxy
disks down to $N({\HI}) = 10^{18}${\cm2}, and in these cases the
disks extend to tens of kpc (\cite{cor93}; \cite{van93}).  
In the M81 group, interactions lead to large $N({\HI})$ in a flattened
distribution well beyond 50~kpc of M81 itself (\cite{yun94}), but some
relatively isolated dwarfs have been found to be extended also
(\cite{hof93}).
As shown in Charlton and Churchill (1996, hereafter CC96), for a disk
with larger scale--height outer regions, the random orientation
cross--section is competitive with that of spherical halos.  
In fact, CC96 argue that after possible biases in the available sample
of {\MgII} absorbers  are taken into account, models can be designed
that are consistent with either a spherical or a thick disk geometry.

Simply because of the cross--section of known {\HI} disks, galaxy
disks {\it must\/} make some contribution to {\MgII} absorption.  
Also, some fraction of absorbing galaxies are diskless (ellipticals),
so at minimum there must be contribution to the absorption by these
two kinematic components.  
Here we address the question of what the dominant contribution is to
{\MgII} absorption.  
Is the larger fraction of the {\MgII} column density from disk
material or from halo material?
Does the answer to this question vary from galaxy to galaxy or from
one line of sight to another within a single galaxy?

Based upon the absorption properties along lines of sight through the
Milky Way and through nearby galaxies, we can begin to infer the
answers to these questions at $z=0$.  
Lines of sight looking out from our special vantage point in the Milky
Way pass mostly through disk material, and from the ratios of various
transitions and their positions in velocity along the line of sight,
the nature of the absorbing clouds can be inferred as {\HI} regions,
{\HII} regions, superbubbles, or high velocity halo clouds
(\cite{sf93}; \cite{fs94}; \cite{sf95}; \cite{wel97}).
As with these many Galactic sight--lines, at higher redshifts there is
considerable kinematic variation in the profiles.
Churchill, Steidel, and Vogt (1996, hereafter CSV96) show that the
variations in the HIRES/Keck profiles of $0.4 < z < 1.0$ {\MgII}
absorbers are not strongly correlated with the impact parameter,
luminosity, or morphological type of the identified absorbing galaxy.
The large scatter in the relationships between absorption and
galaxy properties could be produced by variations due to clumpiness
and discrete structures that lead to variations within the individual
galaxies.

The variations due to clumpy structures will complicate efforts to
extract kinematic information from individual profiles.  
Earlier studies of kinematic signatures demonstrated the basic profile
shapes expected from various spatial and kinematic laws (\cite{wei78};
\cite{lan92}).
Lanzetta and Bowen suggested that a trend might exist where rotating
disk signatures are characteristic of small impact parameter {\MgII}
absorbers, while ``double--horned'' infall profiles arise in the outer
halos of the galaxies.  
The scatter observed in the relationships between impact parameter and
absorption profile properties indicates that the situation is more
complicated (CSV96).
There is hope that a given population of absorbers will give
rise to profiles that reflect the underlying kinematic and spatial
distributions in their host galaxies, but it is important to consider
the effect of stochastic variations on these profiles.  
Prochaska and Wolfe (1997) showed synthetic profiles produced by
clouds selected from a thick exponential rotating disk and found that
this kinematic law is consistent with the properties of high
redshift damped Ly$\alpha$ absorbers.

In this paper we compare the properties of an ensembles of profiles
drawn from various kinematic models (and combinations of kinematic
models) to the ensemble of observed HIRES/Keck profiles of $0.4 < z <
1.0$ {\MgII} absorbers.  
The following four questions motivate our simulations.

\begin{enumerate}

\item Is it possible to extract enough information from a particular
profile to identify it as associated with a particular kind of galaxy,
a line of sight through a particular part of a galaxy, a galaxy
undergoing some particular stage of formation or evolution, etc.?  
To what extent will we be able to resolve the kinematic ambiguities?

\item Can the statistical properties of the observed ensemble be
reproduced by a population of disks and/or halos of galaxies?  
The {\MgII} absorbing galaxies at $0.4 < z < 1.0$ cover all
morphological types and the range of luminosities down to $0.1L^*$.
Are there profiles that require unusual kinematic laws or are all of
them consistent with what we would expect for lines of sight through
the population of galaxies at $z=0$?

\item Can the variety of observed absorption profiles be reproduced
within the context of a single kinematic model, or a weighted
combination of two (such as disk and halo)?  
Can variations in the profiles be due to the expected line of sight
differences through one class of galaxies or are the profiles too
varied?

\item In view of future studies, what role will kinematic analysis of
high resolution absorption profiles of low ionization gas play in
broader studies that also include galaxy properties, Ly$\alpha$, and
higher ionization gas?  
By combining this information, will we be able to extract information
about the conditions along a sight--line through a particular galaxy,
and about the global properties of the galaxy at the same level as is
done through the Galaxy?

\end{enumerate}

The answers to these questions may lead to a characterization of the
population of {\MgII} absorbers at $0.4 < z < 1.0$ in terms of their
kinematic composition.  
In this redshift range the absorbing galaxy can more readily be
identified and studied, providing more leverage on the interpretation
of absorption properties.  
Lines of sight through N--body hydrodynamic simulation boxes (which
also model the ionization states of various metals) show that
different kinematic signatures result from galaxies at different
stages of evolution (\cite{rau97}).
A population of absorbers could be dominated by the kinematics of
material that has just separated from the Hubble flow, material
falling into a galaxy halo, or material in a well--formed disk.
Establishing the contributions of spatial and kinematic components to
{\MgII} profiles at low to intermediate redshift is prerequisite to
interpreting their evolution as the redshift range of observability 
increases through studies in the UV and near--IR (\cite{cwc_uvconf}).

The second section of this paper presents the 26 observed
$0.4 < z < 1.0$ {\MgII} profiles and discusses the selection and
analysis of this sample.  
Even a qualitative examination of these profiles suggests certain
interpretations of their kinematic composition.
Guided by these interpretations and by properties of nearby galaxies
we designed kinematic models.  
The details of the model construction are outlined in
\S\ref{sec:models}, which also describes our statistical comparisons
of the observed and model spectra.
The results of our analysis is given in \S\ref{sec:results}, where we
address which of the kinematic models are formally consistent or
inconsistent with the ensemble of observed spectra.  
In \S\ref{sec:conclude}, we return to the four questions posed in this
introduction and assess the extent to which we can characterize the
kinematic and spatial distribution of the population of $0.4 < z < 1$
{\MgII} absorbers.


\section{HIRES/Keck Observations of $0.4 < z < 1.0$ {\MgII} Absorbers}
\label{sec:observations}

The observed sample for comparison to kinematic models has been
selected from the larger study of {\MgII} absorbers with equivalent
width $W(2796) > 0.3${\AA} (\cite{cwcthesis}; CVC98).
Although the full data set included systems out to $z = 1.8$, for two
reasons we choose to limit this study to the $z < 1$ {\MgII}
absorbers.
First, this sample is unbiased in equivalent width below this redshift
cutoff, while several of the higher redshift systems were selected to
be particularly strong.  
More importantly, there could be an evolution in the kinematic
properties of the global population of galaxies over the larger
redshift interval $0.4 < z < 1.8$.  
The redshift interval of $0.4 < z < 1.0$ covers the lookback time
$\sim 4-9$~Gyr.

\begin{figure}[t]
\plotfiddle{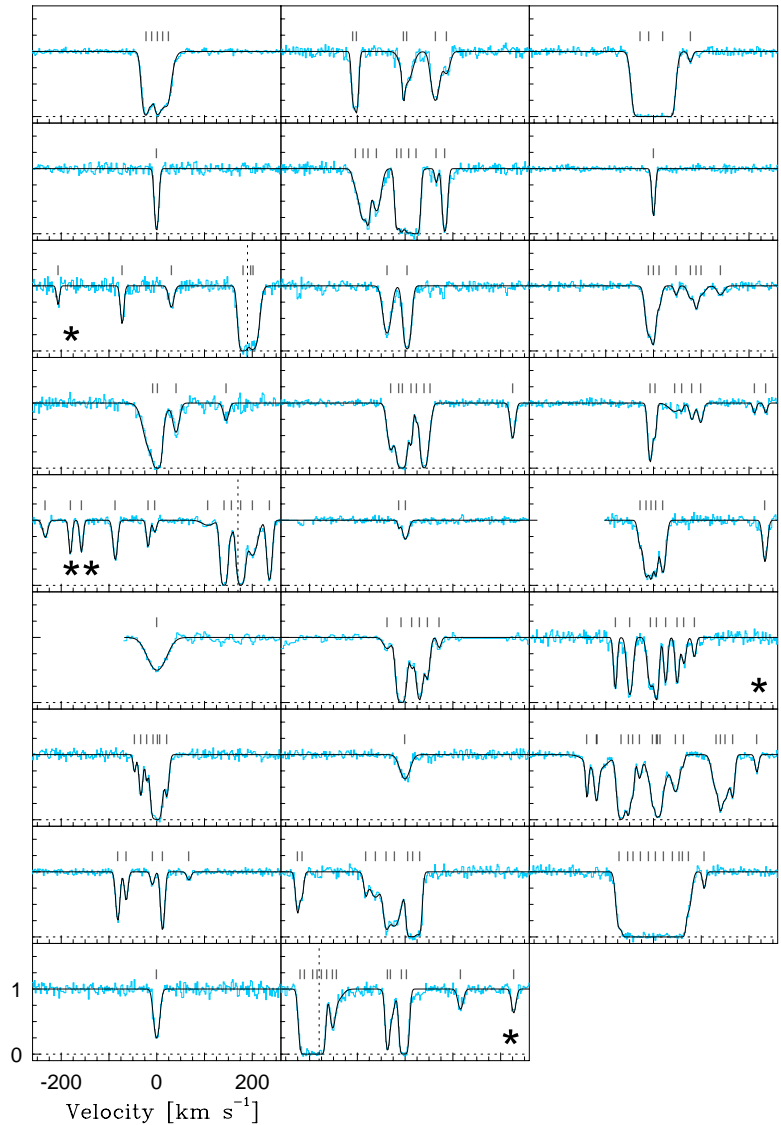}{7in}{0}{65.}{65.}{-200}{50}
\vskip -1.5in
\protect\caption[]
{The observed {\MgII}(2796) profile for $0.4 < z < 1.0$ systems.  The
5$\sigma$ rest--frame equivalent width limits range from 0.007~{\AA}
up to the cutoff of 0.02~{\AA}.  The Voigt profile fits are drawn as
a narrow line superimposed on the data, and the ticks mark the
locations of the centers of the profiles in velocity space.
For those few systems with $v > \pm 260$~{\kms}
clouds, the vertical dotted lines mark the optical depth weighted 
mean, which defines $v=0$~{\kms}.}
\label{fig:data}
\end{figure}

The data were obtained with the HIRES spectrometer (\cite{vog94}) on
the Keck~I telescope and have a spectral resolution of 45,000,
corresponding to 6.6~{\kms} (\cite{cwcthesis}; CVC98).  
In addition to the {\MgIIwaves} doublet, {\MgI}(2853) and several
{\FeII} transitions were observed.
These prove to be important for the study since they allow more
accurate Voigt profile fitting, with a more realistic number of
subcomponents in cases where the {\MgII} doublet is saturated.
So that we have a more uniform sample for the kinematic study we also
eliminate all systems for which the $5 \sigma$ equivalent width
detection limit for {\MgII}(2796) is greater than 0.02{\AA} in the
rest frame.  
In Figure~\ref{fig:data}, we display only the {\MgII}(2796) transition
for each of the 26 systems, with Voigt profile fits determined using
all available {\FeII}, {\MgI}, and {\MgII} transitions.

What follows in this paper is a quantitative comparison of the data to
synthetic spectra from the kinematic models.  
However, visual inspection of the HIRES/Keck {\MgII} profiles in
Figure~\ref{fig:data} already leads us to two simple conclusions about
the kinematics of these objects.
First, many systems are dominated by a strong blended component with a
width of 40--100~{\kms}.  
This range is not consistent with the velocity spread that we would
expect for halo kinematics, whether it be infall, outflow, or random
isotropic motion (for an example of profiles from infall halo
kinematics see Figure~\ref{fig:infallmodels}).
Second, in some profiles there are outlying weaker components spread
over a larger $\Delta v$ of hundreds of {\kms}.  
These large spreads are clearly inconsistent with the kinematics
expected for a rotating disk, even when random vertical motions are
included (see Figure~\ref{fig:diskmodels} for examples of disk
kinematics).
Thus, simply based upon qualitative arguments, {\it it is immediately
apparent that some combination of disk and halo kinematics is
needed to produce the observed ensemble of profiles}.  
This could require either some galaxies that are pure disk galaxies
and others that are pure halo galaxies, or it could imply that some
profiles require both a disk and a halo component in the responsible
galaxy.


\section{The Models and Their Simulated Spectra}
\label{sec:models}

We consider models with rotating disk components and with radial
infall/halo components.  
The model disks are extended-- we assume a radius of $1.5 R(L^*_K)$
where $R(L^*_K) = 35 (L/L^*_K)^{0.15}$~kpc in order that the population of
inclined disks can be consistent with the frequency with which galaxies
within $R(L^*_K)$ of quasar lines of sight produce absorption (CC96).
Halos are assigned a radius of $R(L^*_K)$. 
There are five ``populations'' of kinematic/geometric models that we
present in this study:
1) a single population of pure disk models; 
2) a single population of pure halo models; 
3) a ``two--population'' model, in which 50\% of galaxies are pure disks
and 50\% are pure halos;
4) a single population of ``hybrid'' models, in which every galaxy
has 75\% of its gaseous clouds in a disk and 25\% of its clouds in a
halo (75D/25H); and
5) a single population of ``hybrid'' models, in which
every galaxy has 50\% of its clouds in a disk and 50\% of its
clouds in a halo (50D/50H).

In the course of our study, we investigated two--population models
with various fractions of disk/halo galaxies, but we present only the
50D/50H case for purposes of illustration.
We also investigated other hybrid models, each with a different
fraction of disk/halo clouds.
Again, we show only two of these cases in this paper, but draw upon
the general study for our conclusions.

The full problem of the physical conditions of the clouds is presently
intractable.  
Our goal is simply to test the kinematic distribution of absorbing
clouds and to rule out certain classes of models, if possible (we make
no attempt to diagnose the cloud chemical abundances or
photoionization properties).  
Thus the individual cloud properties are selected from the measured 
distributions of column densities and Doppler parameters
(\cite{cwcthesis}; CVC98).
More precisely, we draw from underlying distributions that produce
profiles consistent with the observed distributions once the simulated
spectra are analyzed in the exact same way as were the data.

The basic simulation procedure is as follows.
For each simulated line of sight we pick a galaxy from the Schechter
luminosity function with $\alpha = -1$.
Its rotation or radial infall velocity is given by the Tully--Fisher
relation.  
Spherical clouds are placed in a spatial distribution and each of
these clouds is assigned a velocity based upon a simple kinematic.
Each cloud is assigned a {\MgII} column density selected from a power
law distribution and a Doppler parameter selected from a Gaussian with
a low--end cutoff.
The ``best'' power law exponent and Doppler parameter spread and
cutoff depends upon the amount of blending in a particular 
model.\footnote{Thus, the simulated spectra for the models must be
analyzed and, in an iterative proceedure, the underlying distributions
are adjusted until the ``observed'' distributions agree with those
from the data.}
Once the model galaxy is defined, we run a line of sight through it at
impact parameter chosen by area weighting considerations and determine
which clouds are intercepted.  
Using the velocities, column densities, and Doppler parameters of
these clouds, we generate simulated spectra.
These spectra are then convolved with the HIRES instrumental profile,
sampled with the pixelization of the HIRES CCD, and degraded to have
signal to noise $S/N$ ratios consistent with the data.
The $S/N$ of any given simulated spectrum is chosen from the
$S/N$ distribution of the data.
We generate the {\FeII} profiles also, assuming thermal scaling of the
Doppler parameters, and applying the observed relation
\begin{equation}
\log N({\FeII}) \simeq \log N({\MgII}) - 0.3 ,
\end{equation}
for the column densities.  Thermal scaling is consistent
with the observed ratios of the {\MgII} and {\FeII}
Doppler parameters (\cite{cwcthesis}; CVC98).
Having followed this procedure we have a set of spectra very similar
to the observations and we can proceed to analyze them in precisely
the same manner.  

The profile fitting procedure used in our analysis is a two step
process.  
The first step is an application of the automated Voigt profile fitter,
{\sc autofit} (\cite{dav97}), which we have modified to account for
broadening due to the instrumental spread function.  
{\sc autofit} was applied to the {\MgII}(2796) transition and this
solution was scaled to give parameters for the Voigt profiles to model
the  {\MgII}(2803) and {\FeII} transitions.
This initial model was refined using {\sc minfit} (\cite{cwcthesis}),
a maximum likelihood least--squares fitter.
{\sc MINFIT} minimizes the $\chi^{2}$ between the data and the Voigt
profile model by adjusting the column densities, Doppler parameters,
and number of Voigt profile components (with the goal of minimizing
the number of these components).

\subsection{Detailed Procedure}

For infall/halo models we place the clouds randomly within the
spherical distribution of radius $R(L_K)$.  
The magnitude of each cloud velocity is set equal to the value given
by the Tully--Fisher relation and its direction is radial
towards the center of the sphere.  
In practice, each cloud is taken to have a radius of 1~kpc, but this
is rather arbitrary since a cloud radius which is smaller would then
require a larger number of clouds per halo such that the number of
intercepted clouds remains conserved at a number consistent with the
data.
Thus we describe each model by its integrated cross sectional area,
$N\sigma _{\rm cl}$, equal to the number of clouds times the cross
section of each.

For disk models there are a larger number of parameters and
assumptions.  
The disk is taken to have a full height of 1~kpc out to a radius of
10~kpc, increasing linearly to a 10~kpc height at its maximum radius
of $1.5 R(L_K)$.
The clouds are rotating differentially with a flat rotation curve with
velocity given by the Tully--Fisher law.  
In addition each cloud is assigned a velocity in the direction
perpendicular to the disk, selected from a Gaussian distribution with
$\sigma = 15$ or $25$~{\kms}.

\begin{figure}[t]
\plotone{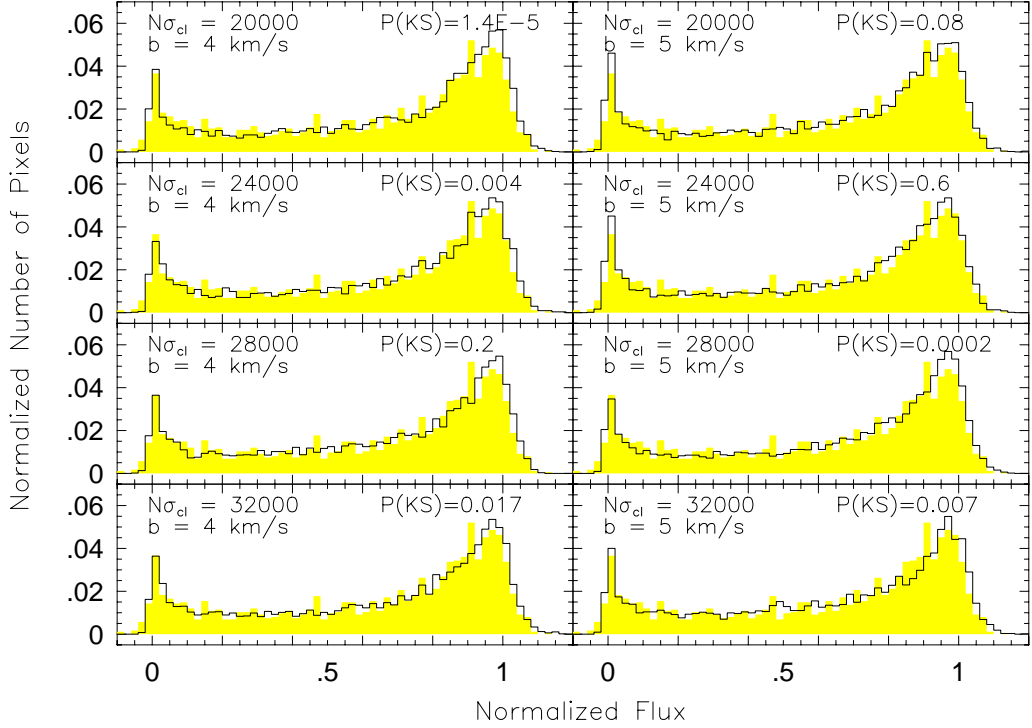}
\protect\caption[]
{Distributions of the number of pixels, in the region of a detected
{\MgII} feature, with a given flux value.   This is a portion of the
grid of 36 models explored for a 50D/50H hybrid model.  The data are
depicted by the shaded region, and the models by the solid
histogram.
Each panel is labelled with the cloud cross section, $N \sigma _{\rm
cl}$, and the mean Doppler parameter of the Gaussian distribution, $b$.
Also shown are the Kolmogornov--Smirnov test probability that the
model fluxes were selected from the same distribution as the data.
For all models, the disk component has a vertical velocity dispersion
$V_z = 25~{\kms}$ and the power--law slope of the input column density
distribution is 1.74. Models with $b=4 (5)~{\kms}$ have a $\sigma(b)=
2 (3)~{\kms}$ and a lower cutoff of 2~{\kms}.  In each vertical
sequence of 4 panels, the gradual increase in the number of saturated
pixels can be recognized as a consequence of the increasing number of
clouds.}
\label{fig:pixdist}
\end{figure}

For a given kinematic model we must find the location in the parameter
space of cloud property distributions that results in spectra that are
consistent with the observed column density and Doppler parameter
distributions.
The parameters that describe the input cloud distributions are the
slope of the {\MgII} column density distribution, and the Gaussian
$\sigma$ and lower cutoff of the Doppler parameter distribution. The
column density distribution has a lower cutoff of $\log N({\MgII}) =
11.3$~{\cm2} for all grid points.  
This is the minimum cloud that could be detected for the highest $S/N$
spectrum in our sample.  
It is not practical to fit all profiles for each grid point, thus we
have devised a test that relies on the flux distribution of all pixels
within absorption features.
In Figure~\ref{fig:pixdist}, this test is illustrated for a small
region of one of our model grids, which includes the cloud properties
and the number of clouds in the galaxy as parameters.  
In fact, because of the large number of pixels in the distributions,
this test provides a sensitive method for refining the grid.  
In order to find a compatible model, it was critically important to
give the model spectra the same distribution of $S/N$ as the data.
From a typical grid of 24--48 choices of parameters (for a given
kinematic model) we considered only models that showed a probability
of more than 1\% by the Kolmogornov--Smirnov test to be drawn from the
same pixel flux distributions as the data.  
Typically, this techniques allowed us to exclude all but 3--5 grid
points, which we then profile fit and conducted further statistical
comparisons.

In Table~\ref{tab:modpars}, we list the parameter choices that yielded
adequate fits for various classes of models, along with their KS
probabilities.
Typically, the effect of blending can be compensated by a steeper
power law for the $N({\MgII})$ distributions, since the lowest flux
pixels are often the result of the combined effects of several lines.
Increasing the vertical velocity dispersion of the disk kinematics
decreases the number of saturated pixels.  
An increase in $N\sigma _{\rm cl}$ can be compensated by a decrease in
the mean Doppler parameter, but ultimately the output Doppler
parameter distribution from fitting must match the observed Doppler
parameter distribution.

\begin{figure}[t]
\plotfiddle{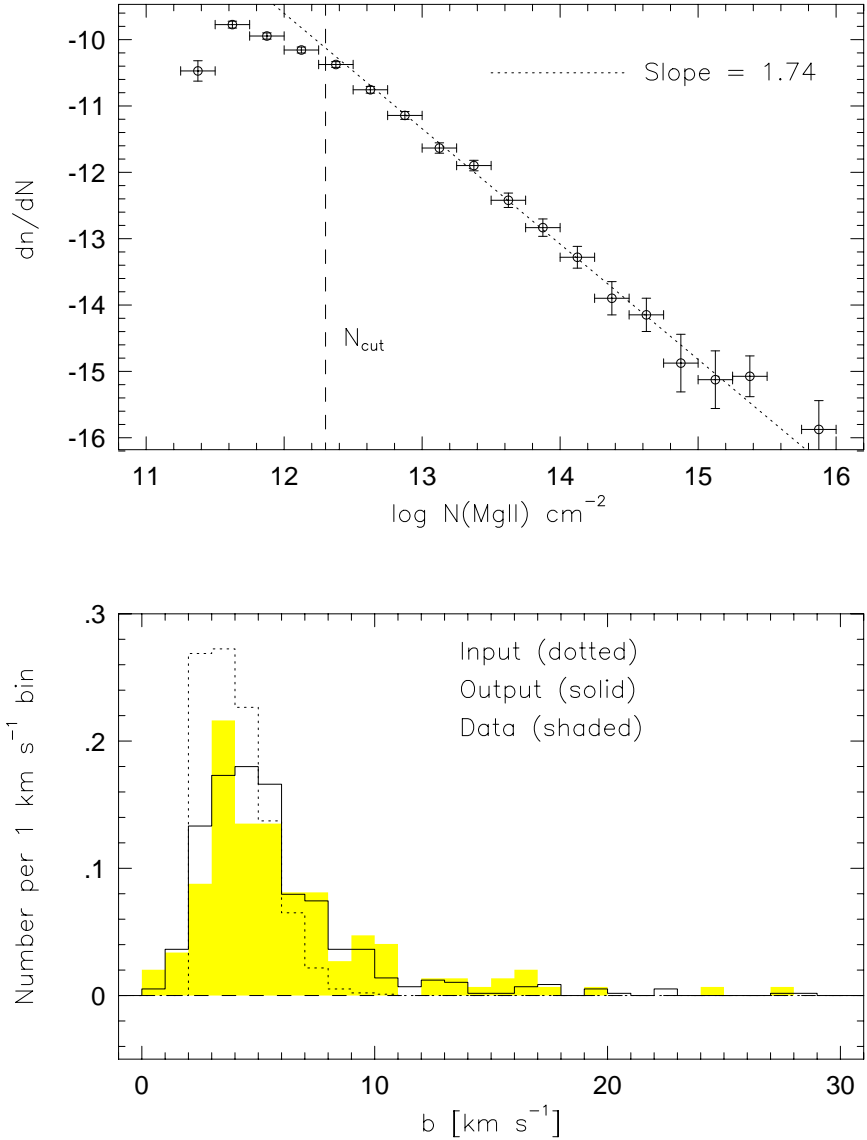}{7in}{0}{65.}{65.}{-200}{50}
\vskip -1.2in
\protect\caption[]
{This figure gives the results for 100 realizations
of a 50D/50H hybrid model.  (upper panel)--- The model column density
distribution output, after Voigt profile fitting.
The model distribution compares well with that of the data,
which were also fit by a power law with slope 1.74 (dotted line) down
to approximately $\log N({\MgII}) = 12.3$.
Except for disk models with small vertical velocity
dispersions, $V_z$, the slope of the column density
distribution is not heavily affected by blending.
However, the lower cutoff is sensitive to the disk velocity
dispersion.
(lower panel)--- The input Doppler parameter distribution (dotted)
is dramatically affected by blending, resulting in a
broader output distribution with a larger mode.  
A KS test shows that the output distribution (solid)
is not inconsistent with that of the data (shaded).}
\label{fig:inoutdists}
\end{figure}

From a maximum likelihood fit to the observed distribution of $0.4
\leq z \leq 1.0$ {\MgII} absorber column densities, we obtain
power--law slope of $\delta = 1.74$ to $n(N) \sim N^{-\delta}$ with
lower cutoff $\log N = 12.5$~{\cm2}.  
In fact, for models with flux distributions consistent with the data,
we found that the slope of the output column density distribution was
similar to the input slope, over a reasonable fitting range.
Therefore an input slope of $\delta = 1.74$ was suitable.
The Doppler parameter distribution of observed data, excluding values
with fractional errors greater than unity, is best fit by a Gaussian
truncated at 2--3~{\kms}, with a peak of 3.5~{\kms} and a standard
deviation of 3--4~{\kms}.
An input Doppler parameter distribution with a mean of 3--4~{\kms}, a
$\sigma$ of 2~{\kms}, and a lower cutoff of 2~{\kms} led to a good fit
of the output distribution with the observed data.
The output distributions of column densities and Doppler parameters
for a consistent hybrid model with 50D/50H contributions are compared
to the input and observed data distributions in
Figure~\ref{fig:inoutdists}.
Based on these considerations we select the following models listed in
Table~\ref{tab:modpars} for further analysis: Pure Disk 2, 75D/25H
Hybrid 2, 50D/50H Hybrid 2, Pure Infall 1, and a two--population mix
in which half the systems are Pure Infall 1 and half are Pure Disk 2

\subsection{Statistical Analysis}

Statistical tests fall into three categories:

(1) {\it Pixel by pixel comparisons}:
These tests involve the distribution of the number of pixels or pixel
pairs satisfying various criteria.
These have the advantage of large number statistics so that F--tests
and KS tests are very sensitive to differences in model distributions.
This is why the pixel flux distribution was judged an effective method
to refine the model grid in the previous section.  
The disadvantage of these types of tests is that they average together
the differences between individual systems, and thus they do not test
whether a model can explain the variation in properties from system to
system.

(2) {\it System by system statistics}: 
The test compare the profile shape and the Voigt profile model cloud
properties system by system.
Examples of these statistics are the various moments of the profiles
of entire systems (weighted by the apparent optical depth) and the
number of clouds per system.  
These test are subject to small number statistics; comparisons can
only be so good as the statistics on the 26 profiles in the observed
sample permit.

(3) {\it Cloud or subfeature statistics}:
These tests are in some ways intermediate between the previous two.
A subfeature is defined as a detected region of a spectrum that is
separated from other detected pixels by at least 2.5 HIRES resolution
elements.
An example of a subfeature statistic is the number of subfeatures
in a given velocity bin.
Cloud properties are parameterized by the Voigt profile fit column
densities and Doppler parameters.
An example of a cloud statistic is the cloud--cloud two point
clustering function.
These statistics combine the individual systems and thus average out
variations between them like (1).  
Although they are subject to small number statistics, there are more
subfeatures and clouds than there are systems, so they are somewhat
better in that sense than (2).

Many plethora of these tests were performed on all models and on the
data.
For presentation we choose a range of tests that best distinguish the
various differences that could exist between our models.
The following battery was selected:
(A) $\Delta V$ of pixel pairs for which both pixels fall in the same
flux bin 0.0--0.2, 0.2--0.4, 0.4--0.6, 0.6--0.8, and 0.8--1.0;
(B) Histogram of number of clouds per system;
(C) Histogram of optical depth weighted velocity widths, $\omega_v$,
for systems, where
\begin{equation}
\omega_v^2 = \int_{v_1}^{v_2} \tau_a(v)\left( v - \left < v \right > 
\right)^2 dv/\int_{v_1}^{v_2} \tau_a(v) dv,
\end{equation}
and $\tau_a$ is the apparent optical depth defined by
\begin{equation}
\tau_a(\lambda) = \ln \left [{I_c(\lambda)} \over {I(\lambda)} 
\right] .
\end{equation}
The zero point of velocity $\left < v \right >$ is defined such that
half of the integrated apparent optical depth lies blueward and half
lies redward of that point;
(D) Histogram of equivalent width for systems;
(E) Histogram of number of subfeatures as a function of velocity;
(F) Two Point Clustering Function, defined as the distribution of
velocity differences for all cloud pairs within single systems, taken
for the full sample of systems.  
The cloud velocities are taken from the Voigt profile fits.

Our formal statistical comparisons between the models and the 
data used the KS test and the F--test.  Neither test is ideal
for comparing distributions with arbitrary shape.  The KS test 
is not sensitive to the tails of the distribution.  The F--test
compares the variance of the distribution, but is not sensitive
to its precise shape.  In order to claim a model is inconsistent
with an observed distribution it is sufficient that only one of 
the two tests gives a small probability that the observed and
model distributions are drawn from the same parent distribution.
It is important to appreciate that if both tests give a relatively
large probability, this does not prove consistency.

Before beginning to discuss the result of these statistical
comparisons for the various models, an important caveat should
be discussed.  
Double galaxies (i.e.~close galaxy pairs) are not treated in our
kinematic models, but some almost certainly do exist in the observed
sample.
The presence of double galaxies can change the statistics
considerably, leading to a much larger number of large velocity clouds
and subfeatures.  
Thus we must be cautious in our interpretations not to hastily rule
out a model because it underpredicts high velocity components.

The make a more quantitative assessment of the double galaxy issue we
consider a second sample (S2) of the observed 26 systems in which we
eliminated from consideration three systems with total velocity
spreads greater than 350~{\kms}.
Those profiles that have been excluded from Sample S2 are marked with
a ``$\ast$'' in Figure~\ref{fig:data}.
It is impossible to unambiguously separate double galaxies and
satellite galaxies from the observed sample, but we have adopted this
fairly extreme approach.
We also separate a fourth system (marked with a ``$\ast \ast$'' in
Figure~\ref{fig:data}) into two systems and include these two instead
in Sample S2.  
The split is made in between the two separate components that are
observed in {\CIVwaves} of the FOS/HST spectrum (\cite{cwc_uvconf};
\cite{chu97}).  
Sample S2 has 24 systems.
We analyzed both the full sample (S1) and Sample S2, and we discuss
both in the following section.


\section{Results of Kinematic Models}
\label{sec:results}

\begin{figure}[t]
\plotfiddle{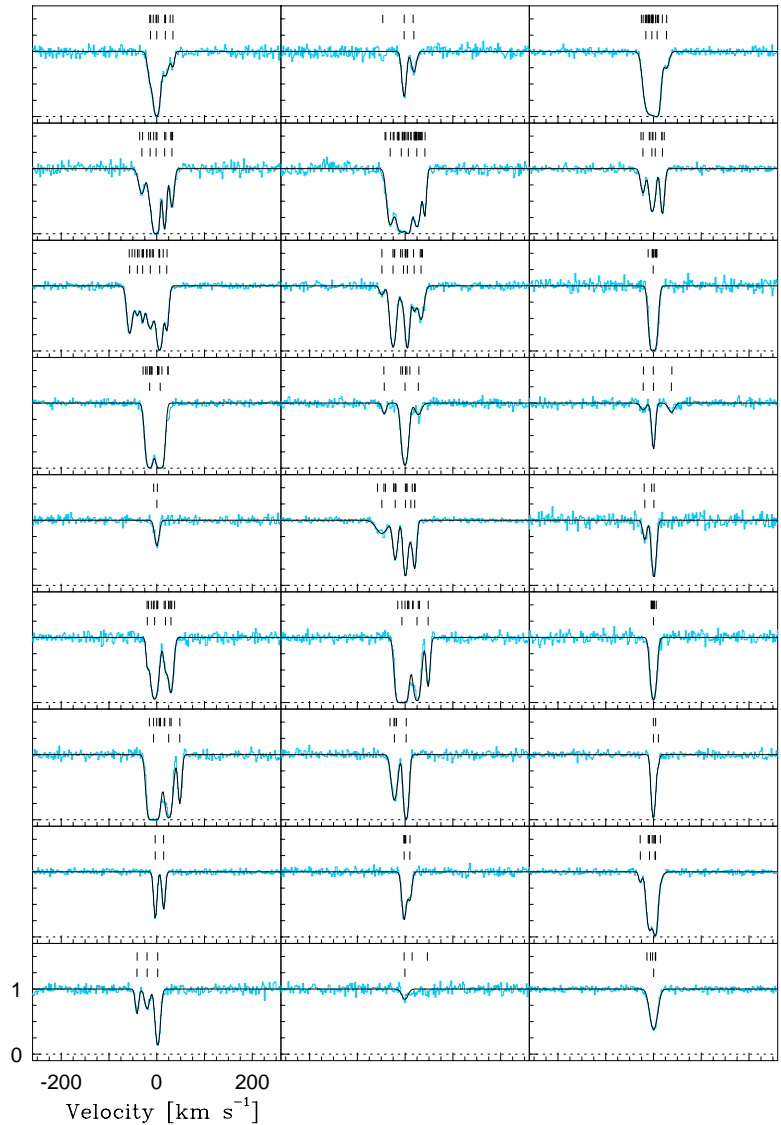}{7in}{0}{65.}{65.}{-200}{50}
\vskip -1.5in
\protect\caption[]
{Twenty seven representative profiles from a pure disk model (Pure
Disk 2 in Table~\ref{tab:modpars}) on the same scale as the data in
Figure~\ref{fig:data}.  
The upper set of ticks represent the velocities at which clouds in the
model were intercepted by the random line of sight.
The lower set are the output components obtained from Voigt profile
model.  More than half of the input clouds are lost to blending in the
disk model, and clearly the Doppler parameter distribution broadens
as a result.}
\label{fig:diskmodels}
\end{figure}

\begin{figure}[t]
\plotfiddle{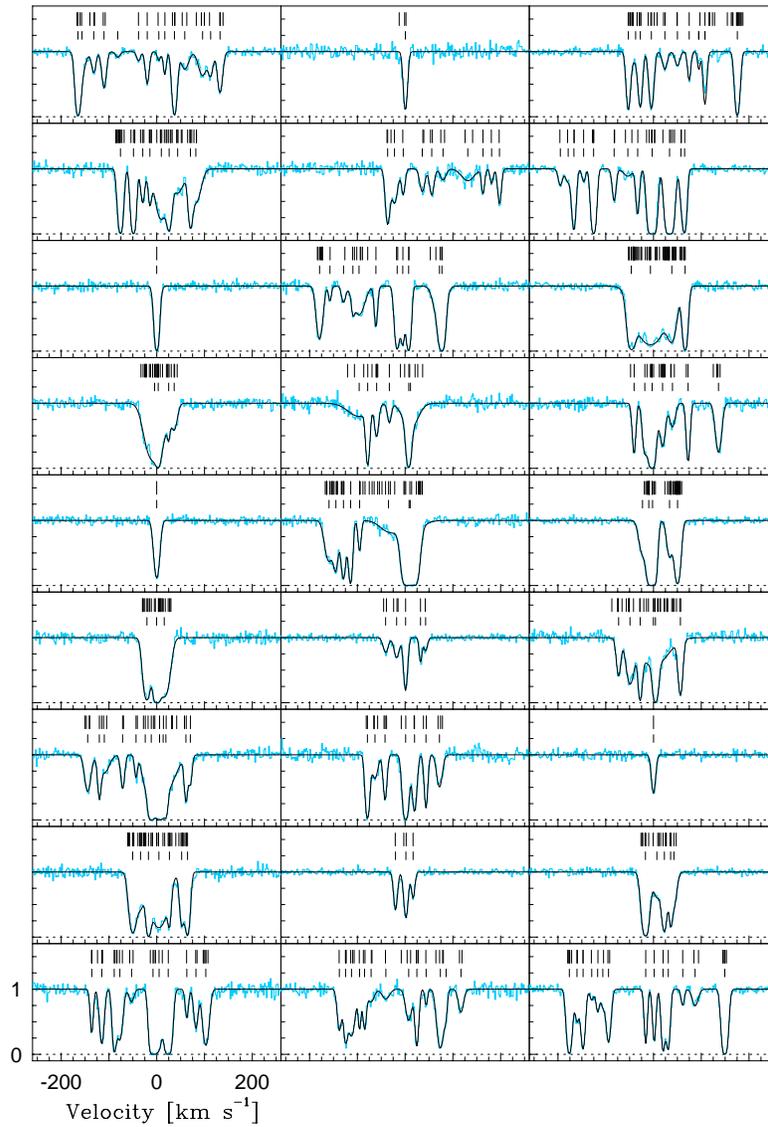}{7in}{0}{65.}{65.}{-200}{50}
\vskip -1.5in
\protect\caption[]
{Same as Figure~\ref{fig:diskmodels}, but for the the infall model
(Pure Infall 1).  Not as large a fraction of components are lost to
blending as in the pure disk models.}
\label{fig:infallmodels}
\end{figure}

\begin{figure}[t]
\plotfiddle{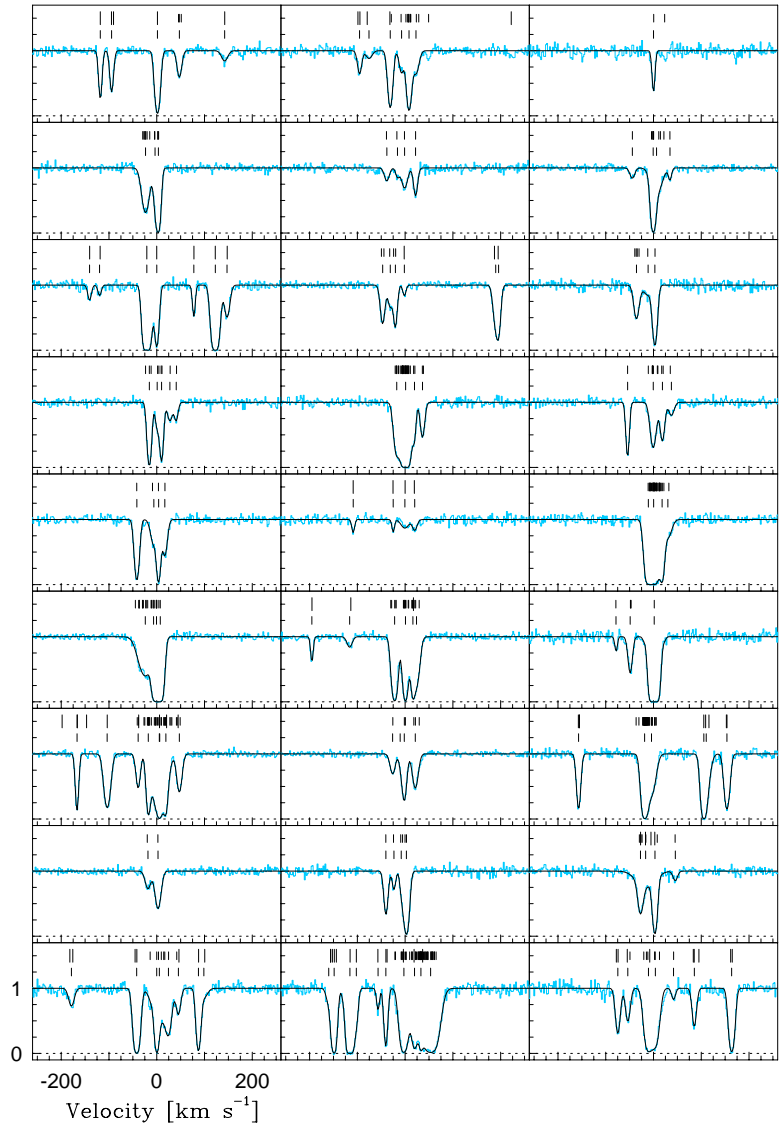}{7in}{0}{65.}{65.}{-200}{50}
\vskip -1.5in
\protect\caption[]
{Same as  Figure~\ref{fig:diskmodels}, but for the hybrid model
(75D/25H Hybrid 2).  
The ``long'' ticks in the upper set mark the input halo clouds and the
``short ticks'' mark the input disk clouds.  In these hybrid models, the
line of sight  often intercepts only disk clouds, though sometimes
only halo clouds are intercepted.  Outlying components in velocity
space (from the disk kinematic center) are sometimes produced by an
occasional halo cloud and this brings the model into better agreement
with the data than a pure disk model.}
\label{fig:hybridA}
\end{figure}

\begin{figure}[t]
\plotfiddle{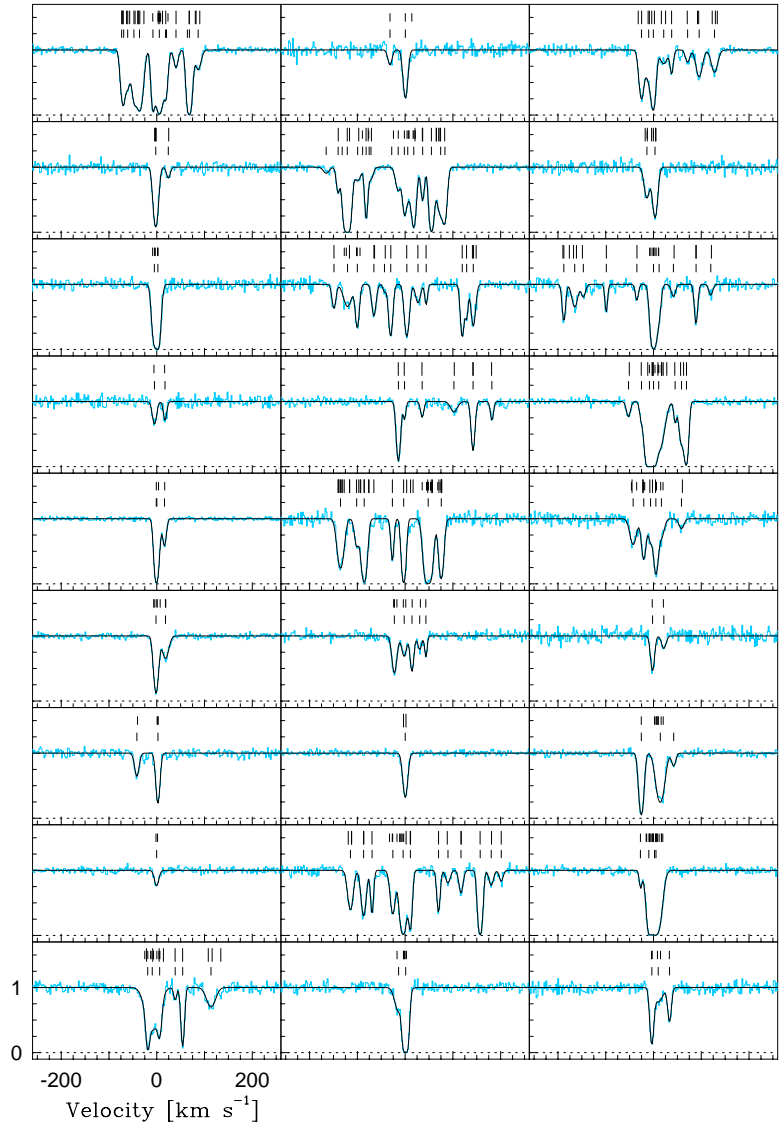}{7in}{0}{65.}{65.}{-200}{50}
\vskip -1.5in
\protect\caption[]
{Same as Figure~\ref{fig:diskmodels}, but for the hybrid model
(50D/50H Hybrid 2).  The ``long'' ticks in the upper set mark the
input halo clouds and the ``short ticks'' mark the disk clouds.  
This high halo to disk fraction results in a larger kinematic spread
than seen for the 75D/25H models (see Figure~\ref{fig:hybridA}).}
\label{fig:hybridB}
\end{figure}

\begin{figure}[t]
\plotfiddle{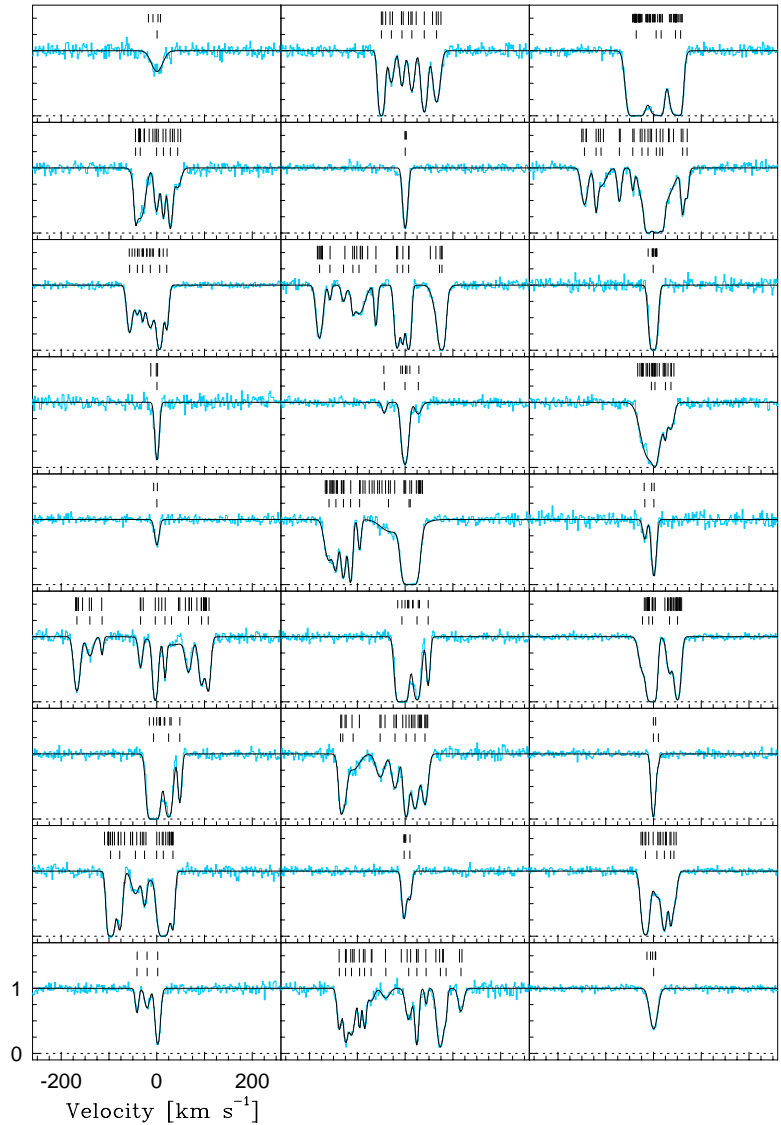}{7in}{0}{65.}{65.}{-200}{50}
\vskip -1.5in
\protect\caption[]
{Same as Figure~\ref{fig:diskmodels}, but for the two--population
model (50\% Pure Disk 1 and 50\% Pure Halo 2).  
This figure allows one to speculate as to which of the profiles
represent disk kinematics and which represent halo kinematics.
The panels with ``long'' ticks in the upper set are from pure
halo models and those with ``short'' ticks are from pure disk models.
In about 1/3 of the cases, the ambiguity is severe and one cannot make
an educated guess as to the component of origin of the profile.}
\label{fig:twopop}
\end{figure}

Sample profiles of the models of various types, selected from the grid
in Table~\ref{tab:modpars}, are presented in
Figures~\ref{fig:diskmodels}--\ref{fig:twopop}.
These models are most consistent with the pixel flux distribution as
well as with the observed column density and Doppler parameter
distributions.
Illustrated are 27 profiles each for: 
the Pure Disk models (Figure~\ref{fig:diskmodels});
the Pure Infall models (Figure~\ref{fig:infallmodels});
the 75D/25H Hybrid models (Figure~\ref{fig:hybridA});
the 50D/50H Hybrid models (Figure~\ref{fig:hybridB});
and,
the 50/50 two--population models (Figure~\ref{fig:twopop}).
The 50/50 two--population models shown in Figure~\ref{fig:twopop} are
drawn from the models presented in Figures~\ref{fig:diskmodels} and
\ref{fig:infallmodels}.
This population illustrates the ambiguity that could exist in
classifying any given observed system.

\subsection{Pixel by Pixel Statistics}

\begin{figure}
\plotfiddle{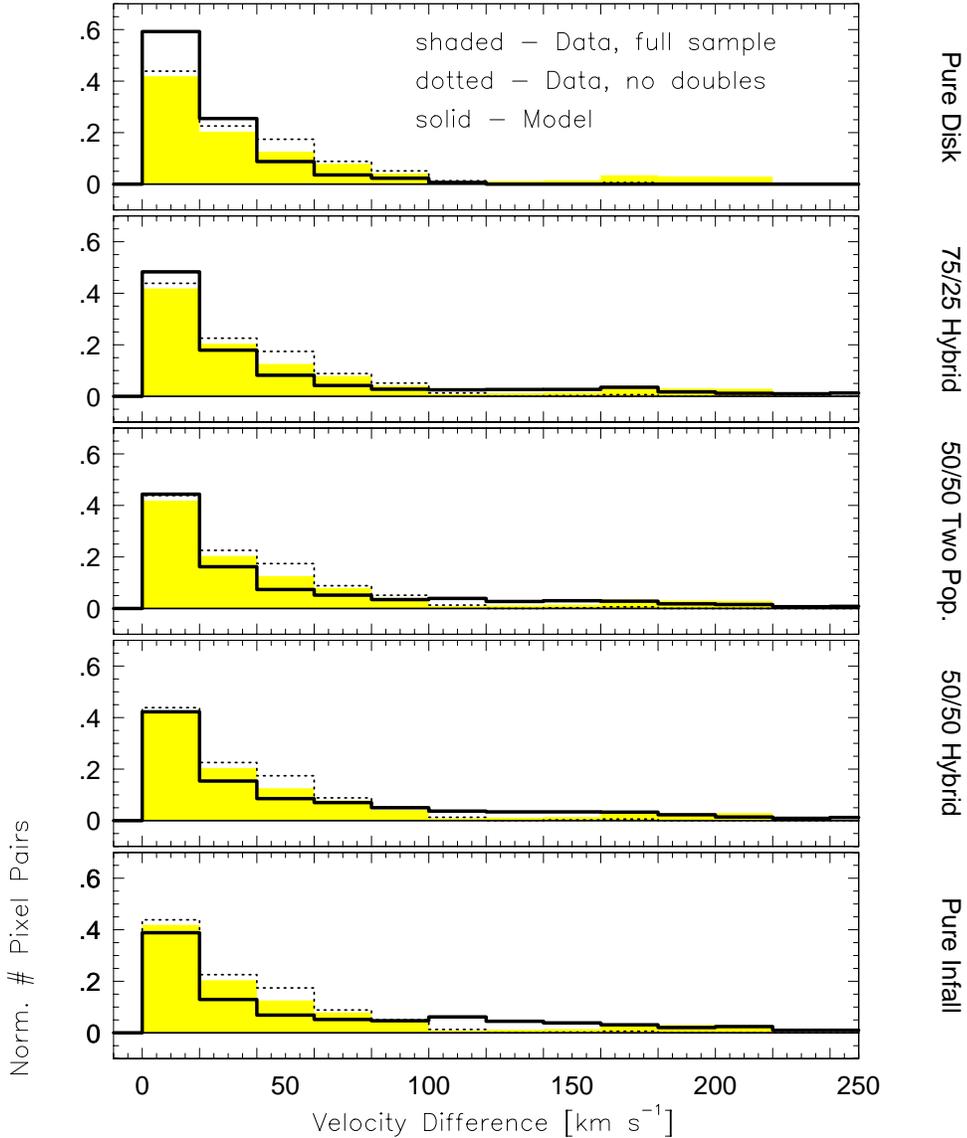}{7in}{0}{65.}{65.}{-200}{100}
\vskip -2.0in
\protect\caption[]
{The distribution of velocity differences between all pairs of pixels
with flux values $< 0.2$ in detected features.  The shaded histogram 
is the distribution for the data (Sample S1).
The dotted histogram is for Sample 2, and illustrates how this
distribution is changed when we attempt to remove all double galaxies.
The solid curves in the five panels are model results, with increasing
infall/halo content from top to bottom.  The 50D/50H hybrid model
produces a similar distribution overall as the two--population mix of
pure disk and pure halo models.  The single population pure disk and
pure halo models are very inconsistent with the data over the full
velocity range.  All models are inconsistent with the data in the
range 20--80~{\kms}, which demonstrates the need for a kinematic
component intermediate between halo and disk.}
\label{fig:velpairs}
\end{figure}

We adapted a procedure developed in Cen \etal (1997) and considered
the pairwise velocity differences for pixels in various flux bins and
within a system.  
We also examined the distributions of velocity relative to the
velocity zero point determined by the apparent optical depth, as
described in \S\ref{sec:models}.  
Since the zero point of the system does not necessarily bear any
relationship to a real kinematic component, we opt to present pixel
velocity differences since they are independent of these
considerations.  
In Figure~\ref{fig:velpairs}, we show the distribution of velocity
differences for all pixel pairs in which both pixels have a flux
between 0.0 and 0.2, i.e. for the saturated pixels.  
In each panel, the model histogram is compared to both observed
samples S1 and S2 (the full sample and one with possible double
galaxies removed).
The pure disk model distribution is quite narrow compared to the data,
and the pure infall model is too broad.  
This is consistent with the visual impression from
Figures~\ref{fig:diskmodels}--\ref{fig:twopop}.
As more disk component is added to a model the distribution gets
narrower and the tail is reduced.  
Two--population models naturally have both the narrow distribution and
the large tail at high velocity differences.

For the 0.0 to 0.2 flux bin, none of the models is in agreement with
the Sample S1 data according to a KS test.  
There are two reasons for this.  
First, the pixel by pixel statistics are quite sensitive tests.  
It is very difficult to tune the parameters so that the distributions
agree to the level where the KS test will not distinguish significant
differences.  
Second, there is a systematic difference between these models and the
data which no amount of fine tuning can alleviate.  
The difference is a deficit of saturated pixel pairs in the bins with
velocity difference 20--80~{\kms}.  
The vertical velocity dispersion for the disk components in displayed
models was $V_z=25$~{\kms}.  
Increasing this value would reduce the inconsistency between the
models and the data, however it would have to be increased
substantially to bring them into agreement.  
At that point we would have a kinematic component that did not bear
much resemblence to a thick disk.  
{\it We conclude that there is a kinematic component contributing to
the observed absorbers that is distinct from a thick disk or a
infall/halo component}.

We also considered the velocity difference distributions for
pixels with flux ranges 0.2--0.4, 0.4--0.6, and 0.6--0.8.  
The same discrepancy present in the 0.0--0.2 bin persists in the
0.2--0.4 bin, however it gradually becomes less extreme.  
Pure infall and pure disk models predict highly discrepant
distributions for the 0.4--0.6 and 0.6--0.8 flux pixels, but the
hybrid 50D/50H and the two--population 50\% pure disk and 50\% pure
halo models, with $V_z=25$~{\kms}, are reasonably consistent
A disk vertical velocity dispersion of 15~{\kms} produces
distributions that are too narrow.

Finally, we consider the effect that double galaxies, that may
exist in the observed sample, would have on these conclusions.  
For this purpose, we also analyzed Sample S2, described at the end of
\S\ref{sec:models}.  
The dotted histogram presented in Figure~\ref{fig:velpairs} 
represents the distribution of velocity differences for saturated
pixels for Sample S2.
Note that compared to Sample S1, the S2 distribution becomes somewhat
narrower and the large velocity difference tail is reduced. 
From this we see that conclusions about specifically which models
provide the best fit are quite sensitive to whether certain particular
observed systems are included in the sample.
However, the large discrepancy between all models and the data in the
20--80~{\kms} bins is pronounced regardless of whether Sample S1 or S2
is used.

\subsection{System By System Statistics}

\subsubsection{Number of Clouds Per System}

\begin{figure}[t]
\plotfiddle{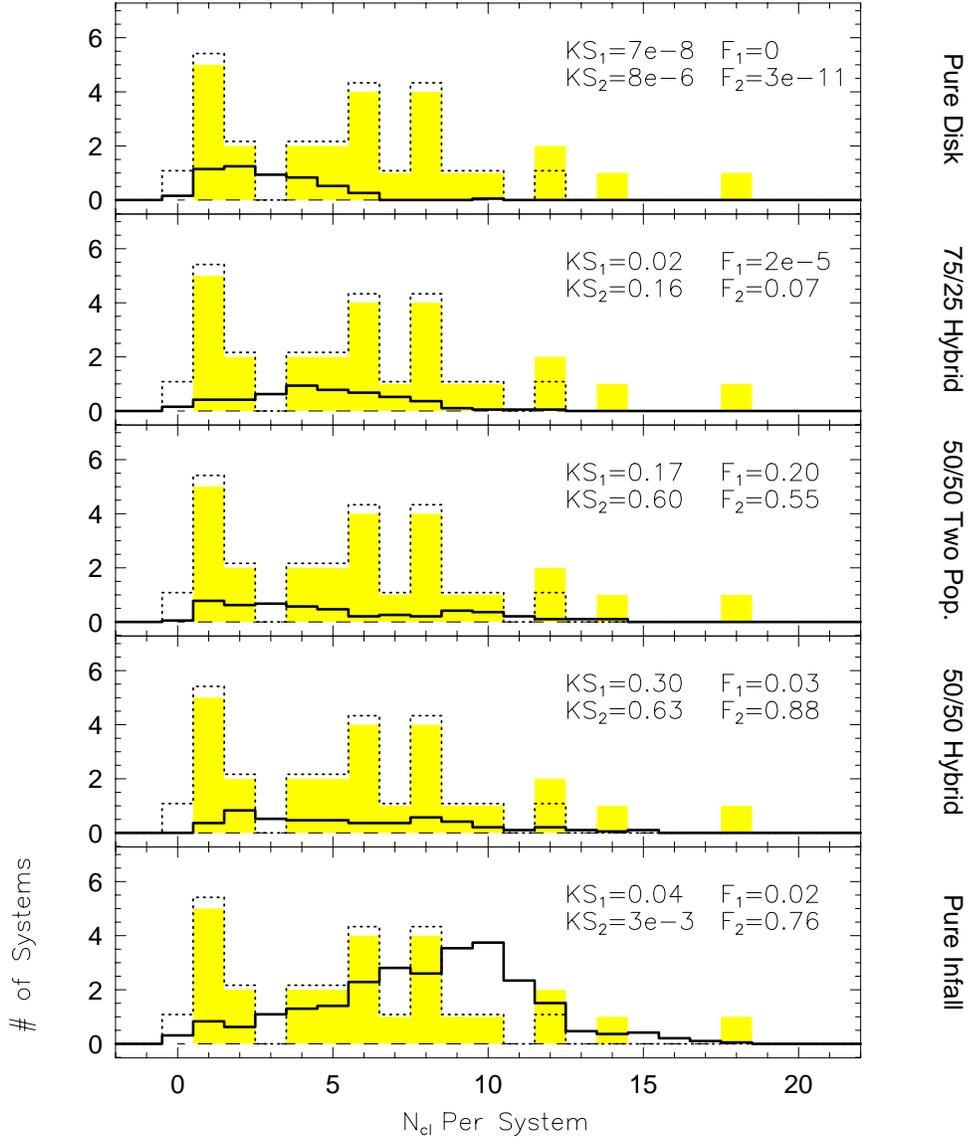}{7in}{0}{65.}{65.}{-200}{0}
\vskip -0.75in
\protect\caption[]
{The distribution of the number of clouds per system, $N_{\rm cl}$,
from the Voigt profile fits.
The shaded histogram is the distribution for Sample S1 of the data.
The dotted histogram is for Sample S2.  
As models become less blended the number of clouds obtained from the
fit increases.
Noted on each panel are the KS and F--test results for comparison
of the model to Samples S1 and S2.
Given the small number statistics, only the pure disk model can be
definitively ruled out by these tests, although the pure infall model
also has low probabilities.}
\label{fig:numclouds}
\end{figure}

As expected, the number of clouds found by the automated profile
fitting procedure decreases as the disk contribution increases, due to
blending.
This is illustrated in Figure~\ref{fig:numclouds}, along with the
observed distribution of the number of clouds per system for Samples
S1 and S2.
The pure disk model is clearly discrepant with both observational
samples, with the KS test yielding a probability less than $7 \times
10^{-8}$ and $8 \times 10^{-6}$ that the model and observational
distributions were drawn from the same distribution.
The hybrid model with 75D/25H contributions does not produce the
outlying points at large $N_{\rm cl}$ in Sample S1, but is consistent
with Sample S2.
The pure infall model is inconsistent with the $N_{\rm cl}$
distribution in Sample S1, and is only marginally consistest by both
the KS and F--tests once the double galaxy systems are removed in
Sample S2.
The two models (hybrid and two--population) with 50/50 
contributions fare best by the combination of tests, each having
greater than 3\% chance of being drawn from the same distribution as
the data for either sample. 

\subsubsection{Velocity Widths}

\begin{figure}[t]
\plotfiddle{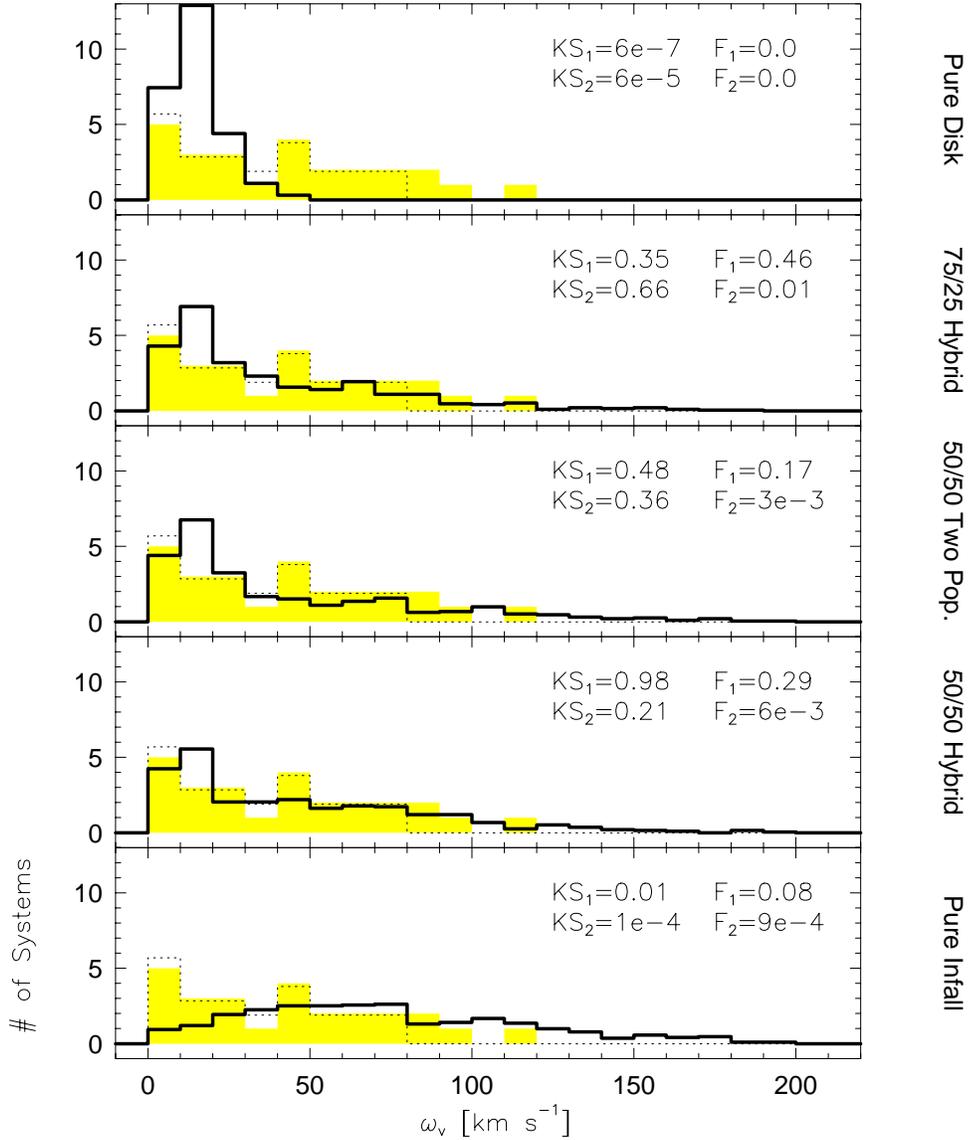}{7in}{0}{65.}{65.}{-200}{0}
\vskip -0.75in
\protect\caption[]
{The velocity width distribution for systems from selected models
compared with Samples S1 (shaded) and S2 (dotted).  The velocity width
distribution depends strongly on the model.  Any of the three
models with both disk and halo kinematics is consistent with
the observed profiles for Sample S1.  All models (except pure
disk) produce an excess of large velocity widths as compared 
to Sample S2, in which an extreme criterion was applied for 
double galaxy removal.}
\label{fig:velmoments}
\end{figure}

In Figure~\ref{fig:velmoments}, the distribution of system velocity
widths for various models is compared to the two observed samples.
The more consistent models, with Sample S1, have between a 50\% and a
75\% disk contribution.  
In all models but the pure disk the high velocity tail is too large
relative to Sample S2.  
However, models could be found that would be consistent with a sample
intermediate between S1 and S2.
Pure infall models, especially, have too large a tail and this
discrepancy becomes larger if we remove the possible double galaxy
contribution to the data.

\subsection{Cloud and Subsystem Statistics}

\subsubsection{Subfeature Velocity Distribution}

\begin{figure}[t]
\plotfiddle{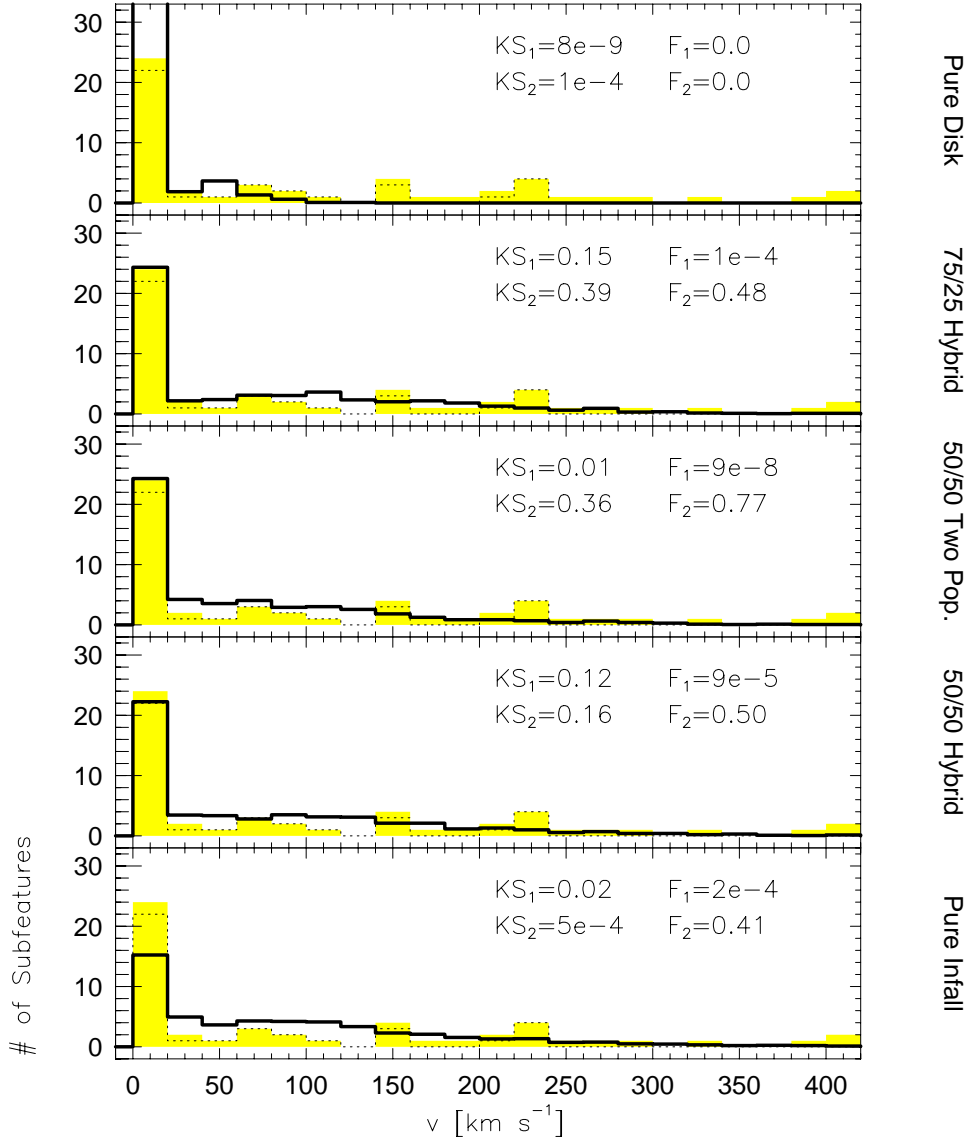}{7in}{0}{65.}{65.}{-200}{0}
\vskip -0.75in
\protect\caption[]
{The velocity distribution for subfeatures in the five
selected models, compared to the two observed samples,
Sample S1 (shaded) and Sample S2 (dotted).
A subfeature is defined as a distinct detected region,
and the velocity of a subfeature is measured relative
to the zero point of the system, defined as the mean of the profile
weighted optical depth.  The tail is much larger for models with
significant halo contribution.  Determining which model is most
consistent with the data is dependent upon which of the observed
samples (with or without suspected double galaxies) is being
compared.}
\label{fig:subfeatures}
\end{figure}

The distributions of central velocities for subfeatures (detections
separated in velocity space by 2.5 HIRES resolution elements or 7
pixels) are presented in Figure~\ref{fig:subfeatures}.
Again, the zero point velocity is defined by the apparent optical
depth technique.  
Pure infall models produce much too large a fraction of subfeatures at
$v > 20$~{\kms}, while pure disk models produce too few.  
The F--test is especially sensitive to the $> 300$~{\kms} observed
data points and thus shows inconsistency between all models and Sample
S1.
All three intermediate models could be consistent with Sample S2,
however.  
The preferred model depends upon which observational sample is being
compared. 
Thus, it is not feasible to finely distinguish between the models
using this test, which is strongly dependent on the high velocity
tail.

\subsubsection{Two Point Clustering Function}

\begin{figure}[t]
\plotone{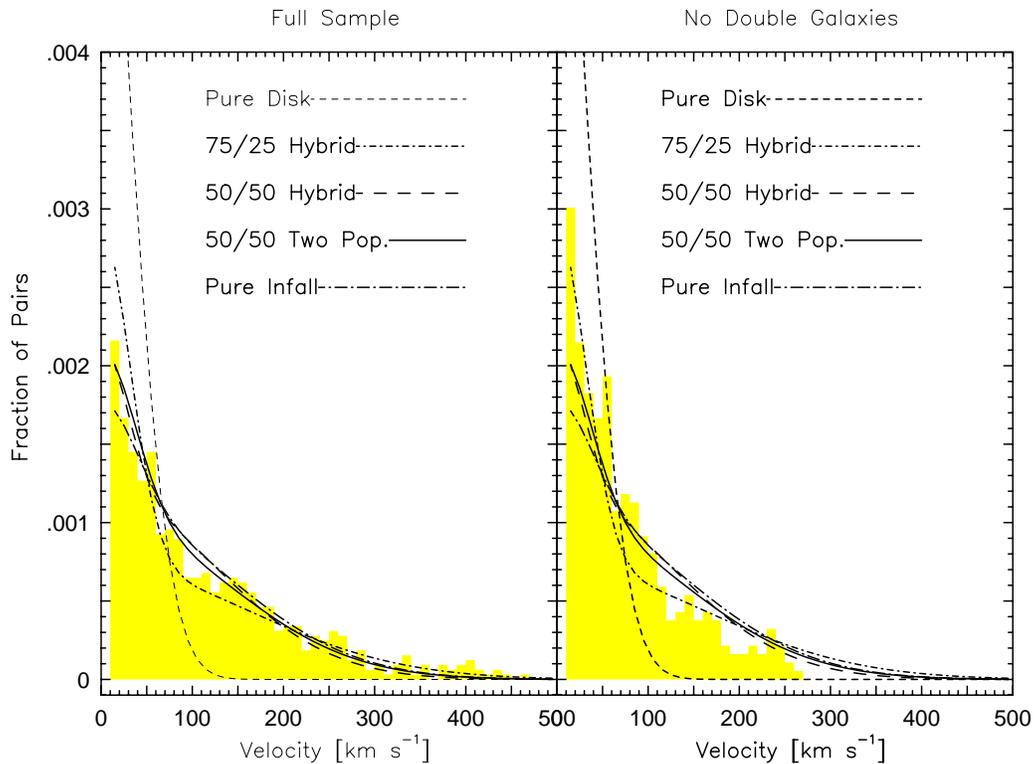}
\protect\caption[]
{The two point clustering function (TPCF) for the five
selected models, compared to Sample S1 (left panel) and to Sample S2
(right panel).  The two observed samples differ dramatically,
indicating that kinematic indicators are often sensitive
to a few systems in a small observational sample.  In either
case, at least two kinematic components seem necessary to
fit the observed distribution.  The vertical velocity dispersion
of the disk component, $V_z$, can be tuned to match the smaller
velocity spread component of the TPCF.}
\label{fig:tpcf}
\end{figure}

The two point clustering function (TPCF) for various models is
illustrated in Figure~\ref{fig:tpcf} by the double Gaussian fits to the
model distibutions.  
There is a systematic shift toward a narrower TPCF as the fraction of
disk contribution is increased.
This is to be compared to the two histograms for the observed Samples
S1 and S2.
For the Full Sample, S1, the best match is to the two--population 50\%
pure disk/ 50\% pure infall model.
The pure disk and 75D/25H hybrid models are too centrally peaked,
while in the 60--140~{\kms} range, the pure halo and the 50D/50H hybrid
models overproduce pairs of clouds as compared to the number observed.
The two--population model combines the appropriate narrow and broad
components, where the narrow component can be well tuned by adjusting
the disk $V_z$.
All models but the pure disk produce too large a tail in the TPCF
compared with Sample S2, but this is very sensitive to just how we
define the observed sample.  
It appears that a large disk contribution is needed for consistency
with this sample.

\subsection{Discussion of Results}

Many more statistical descriptions of the profiles have been
considered than were presented above, but the basic points were
illustrated.  
Although some classes of models can be ruled out, it is in fact quite
difficult to quantify the differences between ensembles of profiles.
A lot of the difference between the models appears as the system by
system characteristics, i.e.~we would like to know whether a given
observed system could be consistent with being drawn from a
distribution of systems produced by a given model.  
With the pixel by pixel statistics the discriminatory power is very
good, so much so that we found a fundamental difference between all
the models and the observed profiles.  
However, even with these tests the information is being considerably
diluted by averaging all of the systems together rather
than looking at them one by one.  
The system by system properties are, as we expected, subject to small
number statistics.
Diluting the rich absorption profile information into a few or even a
single number removes much of the diagnostic leverage available
through direct comparisons of model and observed spectra.

With larger observational samples it may be possible to do somewhat
better in refining a kinematic model grid.  
However, it is more important to note that there is no reason at all
to expect either that all galaxies are the same or that most galaxies
have either pure halo or pure disk contributions to {\MgII}
absorption.  
The models explored here are idealized but they do provide fundamental
information about the kinematic composition of the {\MgII} systems.
We see that both disk--like and halo--like kinematics must contribute
to the profiles, and even that an intermediate kinematic component
appears necessary.  
Roughly equal contributions from halo and disk seem consistent with
the data, but realistically some galaxies, types of galaxies, or
evolutionary stages of galaxies will tend to systematically have more
{\MgII} gas in one component than in the other.


\section{Summary and Conclusion}
\label{sec:conclude}

\begin{figure}[t]
\plotone{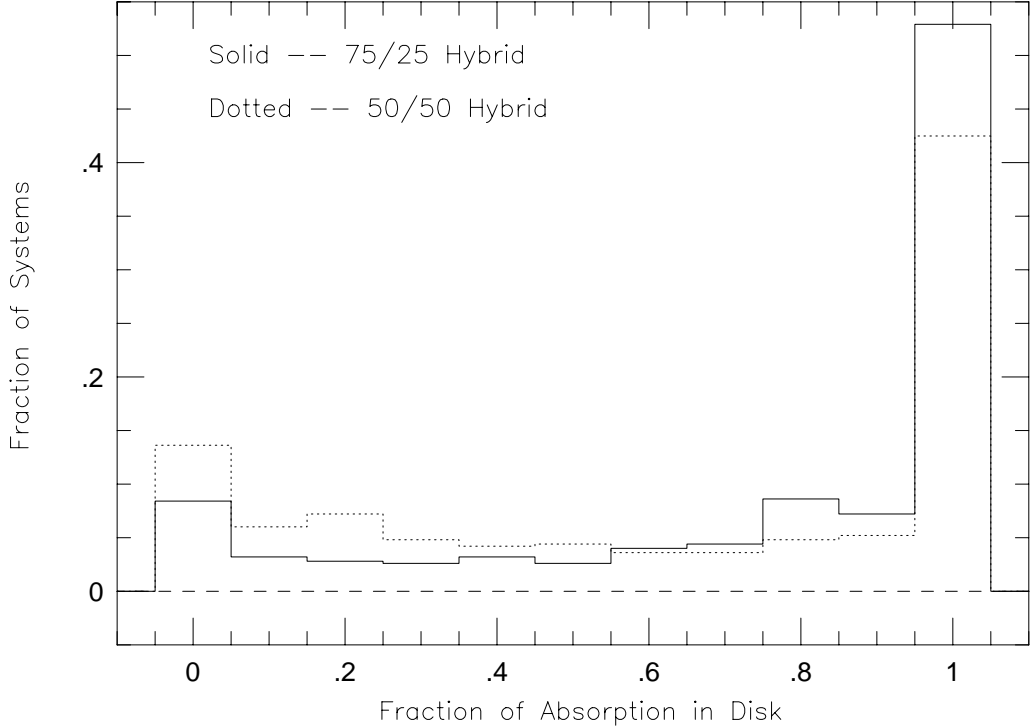}
\protect\caption[]
{For hybrid models, the distribution of the fraction of absorption
contributed by disk clouds.  The solid histogram illustrates that,
for the 75D/25H hybrid model, more than half of the absorbers will
have {\MgII} column densities contributed by disk clouds alone.  
The number is somewhat smaller for the 50D/50H hybrid model,
(dotted histogram).  In both cases, this effect is partially due to
large impact parameter lines of sight that pass only through the
extended disk and not through the halo cloud distribution we have
assumed.  Cross section arguments required our model disks to extend
somewhat beyond our model halos.  The part of the distribution
at $< 0.5$ shows a larger fraction of systems dominated
by halo absorption as the overall halo composition is 
increased from 50--75\%.  Even in these models with significant
(50--75\%) disk contribution, 10--15\% of the systems have absorption
only from the galaxy halo.}
\label{fig:fractions}
\end{figure}

By way of summary, we address the questions raised
in \S\ref{sec:intro} of this paper:
 
It is only possible to extract information about the kinematic
composition of individual galaxies in a statistical sense.  
Some infall/halo models (see Figure~\ref{fig:infallmodels}) could
be mistaken for pure disk models (see Figure~\ref{fig:diskmodels})
if they happen to have few outlying clouds.  
To consider this issue further, a visual inspection of
Figure~\ref{fig:twopop}, in which half of the profiles are from pure
disk and half are from pure halo models, is instructive.
In the context of a particular model we can state a probability that a
given cloud arises in a halo or a disk.
In the hybrid disk/halo models shown in Figures~\ref{fig:hybridA} and 
\ref{fig:hybridB}, we can identify the actual location of origin for
each ``cloud'' (as defined by the Voigt profile models). 
This can then be translated to a probability giving the percentage of
the {\MgII} column density in a system arises from disk material or
from halo material.    
In Figure~\ref{fig:fractions}, we present the distribution of fraction
of disk contribution for lines of sight through the 50D/50H and
through the 75D/25H hybrid models.  
Even though the majority of the {\MgII} absorbing material may be in
the disk, there is a non--negligible probability 
that most of the absorption in a
given system comes from only the halo.  
In some cases we could guess successfully where the absorption arises
from the kinematic signature in the profiles, but in other cases there
are ambiguities that cannot be resolved.

Pure disks absorption or pure halo absorption will not produce sufficient
variety compared to the ensemble of observed profiles. 
However, some individual observed profiles are consistent with being
drawn from one or the other pure kinematic component.
Basically, as could be seen from the qualitative discussion
of the observed profiles in \S\ref{sec:observations}, the disk models
have too few outlying components to be consistent with the data.
The halo models are generally too broad kinematically to be consistent
with the fraction of observed profiles that have spreads of less than
100~{\kms}.

Two different classes of models come close to producing profiles
similar to those observed.  
Which model is most consistent depends on what the contribution is to
the observed profiles by double galaxies, which are not considered in
our kinematic models.  
In both cases, the relatively strong, blended components are produced
by disks and the outlying ones by halos.  
In fact, the two--population models with 50\% pure disk 
and 50\% pure halo model galaxies do produce distributions of pixel velocity
versus flux that are fairly consistent with the data, when averaged
over all systems in the model ensemble.
Although one might expect that the system to system properties would
not match, they are consistent statistically due to the small number
of observed systems.  
The other type of models that produce fair agreement with the data are
the 50D/50H and the 75D/25H hybrid models.
A larger halo component is needed to produce reasonable agreement with
the full observed sample than if suspected double galaxies are
removed.

In all of the models that we have designed there is consistently one
type of disagreement with the observations.
As illustrated in Figure~\ref{fig:velpairs}, the relatively saturated
pixels (with fluxes less than 0.4) are seldom at velocity differences
20--80~{\kms} from the profile center as defined by the apparent
optical depth.  
We conclude that the {\MgII} absorbers have a kinematic component
separate from a halo or thick, rotating disk.  
This separate component must have a characteristic velocity spread
intermediate between halo and disk.
A consistent picture would be one where infalling material is
gradually decelerating and joining the rotation of the disk
(\cite{bd97}).

\subsection{Prospects for Distinguishing Kinematic Components}

This paper has been limited to the interpretation of kinematics of
high resolution spectra of low ionization gas.
With only this information, it is clear that ambiguities remain in
determining which kinematic components are responsible for absorption,
and thus in making global interpretations about galaxy evolution.  
However, there is considerably more information that is or will soon
be available about the properties of the absorbing galaxies and their
gas.  
We conclude by discussing the prospects for studying galaxy formation
and evolution incorporating this additional information.

It is common for the ratio of {\FeII} to {\MgII} to vary significantly
(by more than an order of magnitude) in velocity space across a
system.  
There are two likely causes of this variation: the differing dust
depletion for iron and magnesium, and the relative contributions
of Type Ia versus Type II supernovae (SNe) to the chemical enrichment 
local to the galaxy.  
In the Galactic ISM, iron is depleted by almost one order of magnitude
more than magnesium (\cite{lau96}).  
The relative iron depletion is not as severe in the halo
(\cite{savaraa}), meaning that for a given abundance pattern, the
gas--phase iron to magnesium ratio would be higher in the halo than in
the disk.
Type II SNe produce $\alpha$--elements, such as
magnesium, over short time scales, whereas Type Ia SNe are high iron
producers over a longer timescale.
The abundance patterns of Galactic halo and old disk stars are seen to
have enhanced  $\alpha$--elements and $[\hbox{Fe/H}] < -1$ (\cite{lau96}),
suggesting that in earlier times these stars formed before Type Ia
SNe had significantly influenced the chemical content of our
Galaxy.
The two effects (dust depletion and stellar evolution) work in
opposite directions in their effects on the iron to magnesium
gas--phase abundance ratio, but generally the net result is that iron
is enhanced in disk material.  
If the gas in halos is continuually recycled, then the iron to
magnesium abundance ratio should increase with decreasing redshift as
the Type Ia iron enrichment is cycled into the halo from the disk.
Interestingly, Srianand (1996) found that the equivalent width ratio
of {\FeII}(2382) to {\MgII}(2796) follows this trend, based upon the
low resolution sample of Steidel \& Sargent (1992).
The wild card here, however, is that the UV ionizing spectrum and
possible mechanical ionization sources could be very different in the
disk relative to the halo.
Since ionization corrections are required to infer the abundance
pattern, attempts to exploit this approach will likely be plagued by
uncertainties. 
Similar considerations might allow us to argue whether absorption
arises through material in a dwarf satellite galaxy as opposed to halo
material.
These considerations lead us to suggest that there may promise to
resolve the ambiguity between halo and disk absorption components of
galaxies and to address the level of halo/dnisk interaction in
absorbing galaxies.

\begin{figure}[t]
\plotone{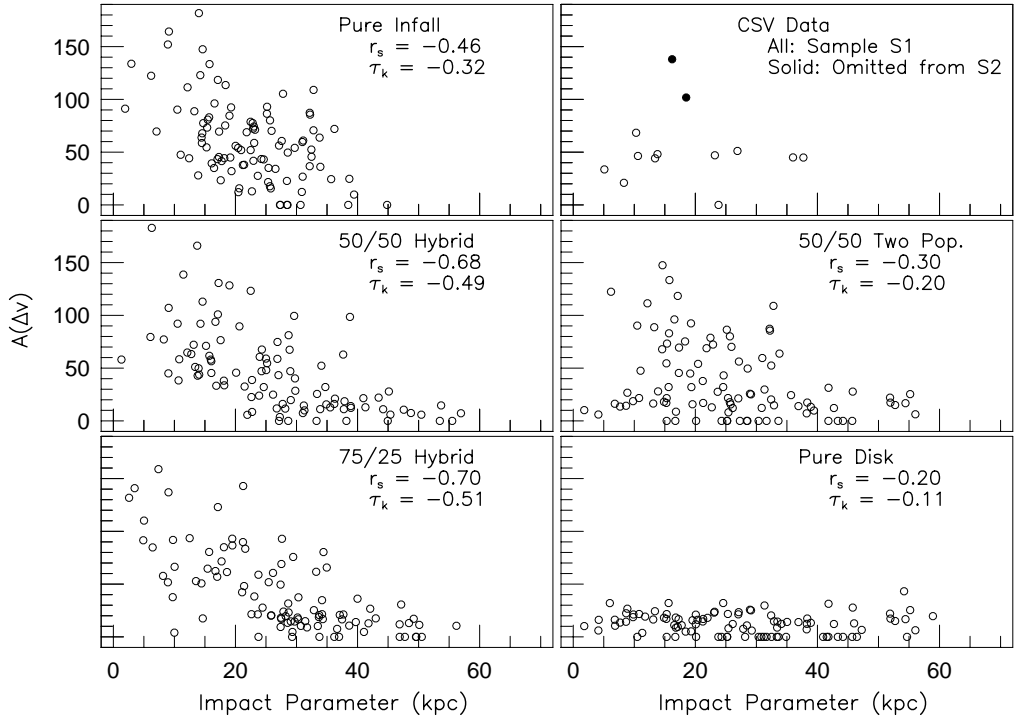}
\protect\caption[]
{The scatter distribution of $A(\Delta v)$ versus the line of sight
impact parameter.  $A(\Delta v)$ is the mean deviation of component
velocities from the median velocity.  Disk models rarely produce
a large spread, while infall models rarely produce a small spread at
small impact parameter.  This represents a technique to distinguish
hybrid models from two--population models.  The upper right panel
shows that the present data set (CSV96) is too small to distinguish
between these models.  Solid circles mark the two systems with
measured impact parameters that are present in Sample S1 but not in
Sample S2 (possible double galaxies removed).}
\label{fig:advimpact}
\end{figure}

It has already been demonstrated that there is a large amount of
scatter in any relationships that might exist between galaxy impact
parameter, luminosity, or color and {\MgII} absorption properties
(CSV96).
Within the context of any given model, we can determine how strong
these relationships should be.
The mean velocity deviation of a subcomponent cloud from the median
velocity,\footnote{See eq.~[1] of CSV96.} $A(\Delta V)$, provides a
good measure of the velocity spread of the system.
In Figure~\ref{fig:advimpact}, we show several models scatter diagrams
of $A(\Delta V)$ of each system versus the impact parameter of
the absorbing galaxy.  
Note the trend for the disk components to have smaller numbers of
clouds and smaller velocity spreads.  
Pure infall models do not have many points in the region of the
diagram at small impact parameter and small $A(\Delta V)$.  
While two--population models can fill in this empty region, hybrid
models cannot because the halo clouds in each galaxy combine with the
disk clouds to create a larger $A(\Delta V)$.  
Identification of the absorbing galaxies for a larger sample of
{\MgII} absorbers should allow us to distinguish between
two--population and hybrid disk/halo models.  
More simply stated, it will be possible to determine whether or not it
is common for disk galaxies to have little {\MgII} absorbing material
in their halos at small impact parameter.

The distribution of the high ionization gas associated with the
population of $0.4 <z <1$ {\MgII} absorbers should also be incorporated
into an overall kinematic analysis of these systems.  
The high and low ionization gas are quite possibly distributed in
distinct kinematic structures or in different phases of a multi--phase
medium (\cite{gir94}).  
There is a large variety in the relationship between high and low
ionization gas in the population of $z>2.5$ damped Ly$\alpha$
absorbers, i.e.~sometimes they appear to trace the same kinematic
components and sometimes they do not (\cite{lu96}).  
When they do not, the high ionization gas tends to have a larger
velocity spread.  
In $z < 1$ {\MgII} absorbers, it has not been possible to assess
whether high and low ionization gas are kinematically distinct because
of a lack of high resolution data for the high ionization resonance
doublets ({\CIV}, {\SiIV}, {\NV}, {\OVI}) that lie in the UV. 
With HST/STIS it is possible to collect these data and to compare, for
example, the kinematic components of low $z$ {\CIV} profiles to those at
high $z$.  
At high $z$, according to simulations, these profiles show indications
of gas separating from the Hubble flow and falling into halos
(\cite{rau97}).
At lower $z$ will we also see evidence of further kinematic settling,
in the form of a disk--like component in the {\CIV} profiles, as we
have shown must be present in some of the {\MgII} absorbers?

From a kinematic analysis of {\MgII} profiles we concluded (point 1
above) fairly negatively about the prospects to describe an individual
{\MgII} system in terms of its specific kinematic components.
Although it is not possible to unambiguously diagnose the nature of
an individual {\MgII} absorber, we can sometimes find a large
probability that a particular subcomponent is produced by halo
material or by disk material.
In the future it will be possible to incorporate into kinematic
interpretation the additional information on the ratios of {\MgII} to
{\FeII}, on the absorbing galaxy impact parameter, orientation and
morphology as well as those of other galaxies in the field, and on the
high ionization gas and Ly$\alpha$ absorption.
With this information, the prospects for interpretation of the
individual galaxy properties are significantly improved and a more
detailed study of galaxy evolution is possible.

\acknowledgments

This work was supported by
the National Science Foundation under Grants AST-9529242 and
ASST-9617185.  C.W.C. acknowledges support by the Eberly College 
of Science Distinguished Postdoctoral Fellowship at Penn State.
We gratefully acknowledge Lester Chou for his expert assistance
in preparing figures for this paper.  Special thanks also to Rajib
Ganguly for valuable technical and interpretational suggestions
at various stages of this project.  We thank S. Vogt for his
work building the HIRES spectrograph.  Romeel Dav\'e generously
provided his {\sc autofit} code.  We have also benefit from
conversations with many colleagues, especially R. Cen,
M. Dickinson, K. Lanzetta, H. Mo, J. Prochaska, C. Steidel, and 
A. Wolfe.  


\vfill\eject

\begin{deluxetable}{lcccccccc}
\tablecaption{Selected Model Parameters}
\tablehead
{
\colhead{Model} & 
\colhead{$N\sigma _{\rm cl}$(disk)} & 
\colhead{$N\sigma _{\rm cl}$(halo)} &
\colhead{$\delta$} &
\colhead{$V_z$} &
\colhead{$\left< b \right> $} &
\colhead{$\sigma(b)$} &
\colhead{$b_{\rm cut}$}&
\colhead{$P(KS)$} \\
\colhead{(1)} & 
\colhead{(2)} & 
\colhead{(3)} &
\colhead{(4)} &
\colhead{(5)} &
\colhead{(6)} &
\colhead{(7)} &
\colhead{(8)}&
\colhead{(9)}
}
\startdata
Pure Infall 1 & 0 & 28000 & 1.74 & -- & 3 & 2 & 2 & 0.15 \nl
Pure Infall 2 & 0 & 24000 & 1.74 & -- & 4 & 2 & 2 & 0.22 \nl
50D/50H Hybrid 1 & 14000 & 14000 & 1.74 & 15 & 3 & 2 & 2 & 0.22 \nl       
50D/50H Hybrid 2 & 14000 & 14000 & 1.74 & 25 & 3 & 2 & 2 & 0.21 \nl
50D/50H Hybrid 3 & 16000 & 16000 & 1.74 & 15 & 4 & 2 & 2 & 0.11 \nl
50D/50H Hybrid 4 & 16000 & 16000 & 1.74 & 25 & 4 & 2 & 2 & 0.13 \nl
50D/50H Hybrid 5 & 16000 & 16000 & 1.74 & 15 & 3 & 2 & 2 & 0.09 \nl
50D/50H Hybrid 6 & 16000 & 16000 & 1.74 & 25 & 3 & 2 & 2 & 0.38 \nl
50D/50H Hybrid 7 & 12000 & 12000 & 1.74 & 15 & 5 & 3 & 2 & 0.41 \nl
50D/50H Hybrid 8 & 12000 & 12000 & 1.74 & 25 & 5 & 3 & 2 & 0.06 \nl
75D/25H Hybrid 1 & 21000 & 7000 & 1.74 & 15 & 4 & 2 & 2 & 0.07 \nl
75D/25H Hybrid 2 & 21000 & 7000 & 1.74 & 25 & 4 & 2 & 2 & 0.20 \nl
75D/25H Hybrid 3 & 18000 & 6000 & 1.74 & 15 & 5 & 3 & 2 & 0.12 \nl
75D/25H Hybrid 4 & 18000 & 6000 & 1.74 & 25 & 5 & 3 & 2 & 0.60 \nl
Pure Disk 1 & 14000 & 0 & 1.74 & 15 & 4 & 2 & 2 & 0.12 \nl
Pure Disk 2 & 14000 & 0 & 1.74 & 25 & 4 & 2 & 2 & 0.03 \nl
Pure Disk 3 & 12000 & 0 & 1.84 & 15 & 5 & 3 & 2 & 0.10 \nl
Pure Disk 4 & 12000 & 0 & 1.84 & 25 & 5 & 3 & 2 & 0.11 \nl
Pure Disk 5 & 12000 & 0 & 1.84 & 15 & 4 & 2 & 2 & 0.15 \nl
Pure Disk 6 & 12000 & 0 & 1.84 & 25 & 5 & 3 & 2 & 0.18 \nl
Pure Disk 7 & 10000 & 0 & 1.84 & 15 & 5 & 3 & 2 & 0.10 \nl
Pure Disk 8 & 10000 & 0 & 1.84 & 25 & 5 & 3 & 2 & 0.02 \nl
\enddata
\tablecomments{
(1) The model name used throughout this paper. 
(2) The effective cloud cross section in kpc$^2$ for the disk component.
(3) The effective cloud cross section in kpc$^2$ for the halo component.
(4) The column density distribution input power law slope.
(5) The cloud vertical velocity dispersion in the disk.
(6) The input mean Doppler parameter.
(7) The Gaussian width of the Doppler parameter distribution.
(8) The lower cutoff in the Doppler parameter distribution.
(9) The KS probability that the flux distribution of the model
profiles is not inconsistent with that of the data.}
\label{tab:modpars}
\end{deluxetable}

\end{document}